\documentclass[paper]{ieice}
\usepackage[T1]{fontenc}
\usepackage{times}
\usepackage{booktabs}
\usepackage{multirow}

\usepackage{tikz}

\usepackage{booktabs}
\usepackage{graphicx}
\usepackage{nicefrac}
\usepackage[T1]{fontenc}
\usepackage{soul}
\usepackage{xspace}
\usepackage{color}
\usepackage{url}
\usepackage{booktabs}
\usepackage{multirow}
\usepackage{listliketab}
\usepackage{amssymb}
\usepackage{algorithmic}
\usepackage{booktabs}
\usepackage{comment}
\usepackage{amsmath}
\usepackage{bbding}
\usepackage{listings}
\usepackage{float}
\usepackage{tabularx}
\usepackage{cite}
\usepackage{rotating}
\usepackage{tcolorbox}
\usepackage{ulem}
\usepackage{censor}
\usepackage{tikz}
\usepackage{amssymb} 
\usepackage{amsfonts}
\usepackage{graphicx,subfigure}
\usepackage{xparse}
\usepackage[colorinlistoftodos]{todonotes}
\usepackage{ulem}
\usepackage{amssymb}
\usepackage{lmodern}
\usepackage{mathtools, nccmath}
\usepackage{venndiagram}
\usepackage[symbol]{footmisc}

\usepackage{balance}
\usepackage{esvect}
\usepackage{xurl}
\usepackage[colorinlistoftodos]{todonotes}

\newcommand{\RQOne}{\textbf{RQ$_1$: To what degree do developers request collaborations when posting patch linkages?}}
\newcommand{\RqTwo}{\textbf{RQ$_2$: How likely will collaborations occur after patch linkages are posted?}}
\newcommand{\RqThree}{\textbf{RQ$_3$: What are the kinds of cross-patch collaboration activities?}}

\newlength\WIDTHOFBAR
\def\blackwhitebar#1{%%
  #1 {\color{black!100}\rule{#1cm}{8pt}}{\color{black!30}\rule{\WIDTHOFBAR - #1 cm}{8pt}}}

	\def\mybar#1{%%
  {\color{black}\rule{#1cm}{8pt}}}

\usepackage[english]{babel}

\makeatletter
\def\@xfootnote[#1]{%
  \protected@xdef\@thefnmark{#1}%
  \@footnotemark\@footnotetext}
\makeatother

\field{D}
\vol{103}
\no{1}
\SpecialSection{Empirical Software Engineering}

\title{An Exploration of Cross-Patch Collaborations via \\ Patch Linkage in OpenStack}

\authorlist{%
\authorentry{Dong Wang}{n}{kyushu}\MembershipNumber{}
\authorentry{Patanamon Thongtanunam}{n}{mel}\MembershipNumber{}
\authorentry{Raula Gaikovina Kula}{n}{naist}\MembershipNumber{}
\authorentry{Kenichi MATSUMOTO}{f}{naist}\MembershipNumber{8925358}
}
\affiliate[kyushu]{Kyushu University, Japan.
}
\affiliate[mel]{The University of Melbourne, Australia.
}
\affiliate[naist]{Nara Institute of Science and Technology, Japan.
}

\begin{document}
\maketitle
\begin{summary}
\noindent Contemporary development projects benefit from code review as it improves the quality of a project.
Large ecosystems of inter-dependent projects like OpenStack generate a large number of reviews, which poses new challenges for collaboration (improving patches, fixing defects).
Review tools allow developers to link between patches, to indicate patch dependency, competing solutions, or provide broader context.
We hypothesize that such patch linkage may also simulate cross-collaboration.

With a case study of OpenStack, we take a first step to explore collaborations that occur after a patch linkage was posted between two patches (i.e., \texttt{cross-patch collaboration}).
Our empirical results show that although patch linkage that requests collaboration is relatively less prevalent, the probability of collaboration is relatively higher.
Interestingly, the results also show that collaborative contributions via patch linkage are non-trivial, i.e, contributions can affect the review outcome (such as voting) or even improve the patch (i.e., revising).
This work opens up future directions to understand barriers and opportunities related to this new kind of collaboration, that assists with code review and development tasks in large ecosystems.
\end{summary}
\begin{keywords}
Collaboration, Human Aspects, Code Review
\end{keywords}

\section{Introduction}
\label{intro}
 Software development teams nowadays benefit from  online code review tools (e.g., Gerrit, Codestriker, and ReviewBoard) to effectively inspect patches and improve the code quality of their software systems, 
while enabling the teams to perform asynchronous code reviews that are more lightweight and flexible~\cite{WANG2021111009, Google_2018}.
On the other hand, a large number of code reviews are being performed by software teams as new patches (i.e., a set of code changes) frequently occur in a contemporary code review setting~\cite{FSE2013_Rigby}. 
For example, the 2018 OpenStack User Survey report\footnote[1]{https://www.openstack.org/user-survey/2018-user-survey-report/} showed that about 70,000 patches were reviewed, with an average of 182 code reviews changes per day.
Such the large number of code reviews potentially poses a new challenge for collaboration (e.g., improving the patch, fixing defects) during code reviews and development tasks.

More specifically, recent studies have highlighted evidence of why developers should collaborate across code review tasks.
Zhang
et al.~\cite{ICSME18_pull} found that redundant patches (i.e., patches that address the same task or problem) are often submitted for a review in software projects hosted in GitHub.
Ebert et al.~\cite{ebert2019confusion} observed that the inclusion of more people in the code review increases their awareness of the code change, i.e., confusion resolution contributes to knowledge sharing.
Recently, Wang et al.~\cite{WANG_emse} observed that developers are likely to share links during review discussions with several intentions to fulfill information needs.
Meanwhile, Hirao et al.~\cite{hirao2019fse} shed light that the patch linkage (i.e., posting a patch link to another patch) is used to indicate patch dependency, competing solutions, or provide broader context.
As recent work has shown that patch linkage can increase the awareness of the related patches, we further investigate to what extent developers collaborate across these linked patches.

\begin{figure}[t]
\centering
\includegraphics[width=\linewidth]{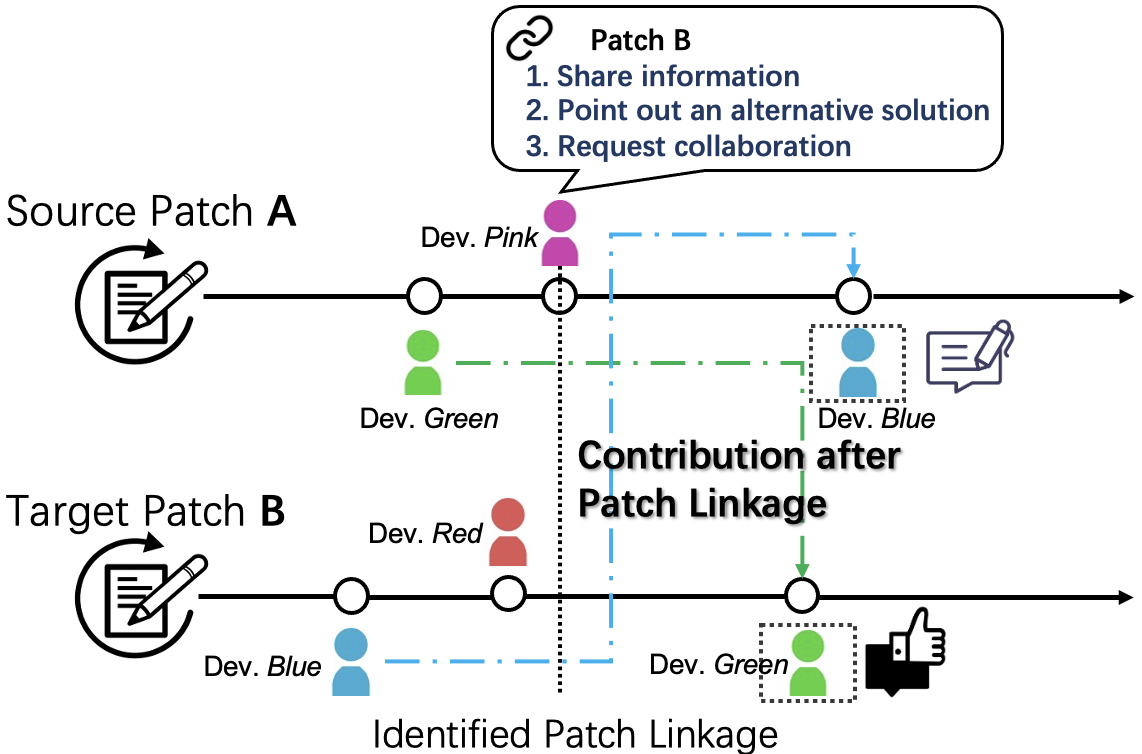}
\caption{A conceptual illustration that describes (1) a linkage between two patches is identified and (2) a collaboration activity happens where a developer on one patch contributes to the review of the other patch.}
\label{fig:illustration}
\end{figure}

\begin{figure*}[t]
    \centering
    \subfigure[In the example, one developer left a comment in Patch \texttt{211019}, in order to request a collaboration with another patch \texttt{209612}.]{\includegraphics[width=.7\linewidth]{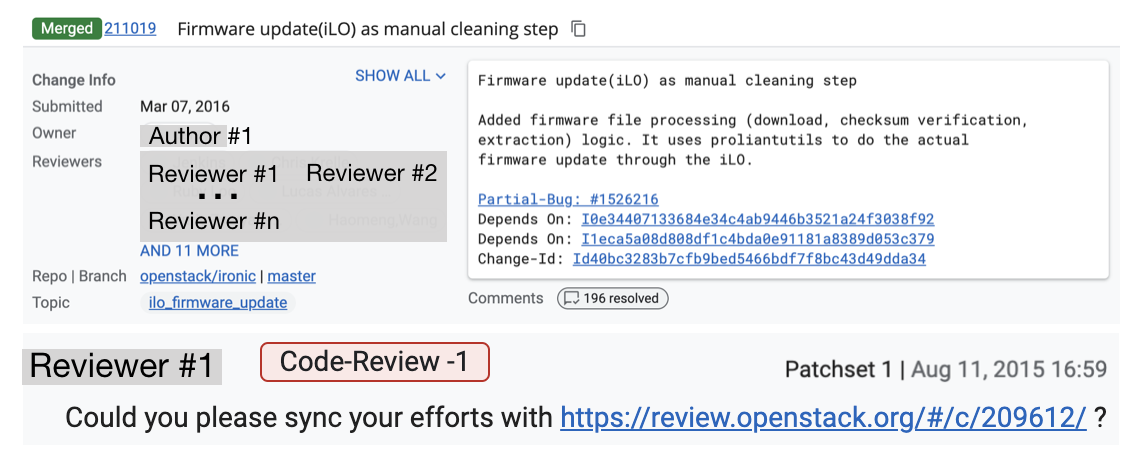}}
    \subfigure[In Patch \texttt{209612}, the author and one reviewer from Patch \texttt{211019} provided code comments. ]{\includegraphics[width=.7\linewidth]{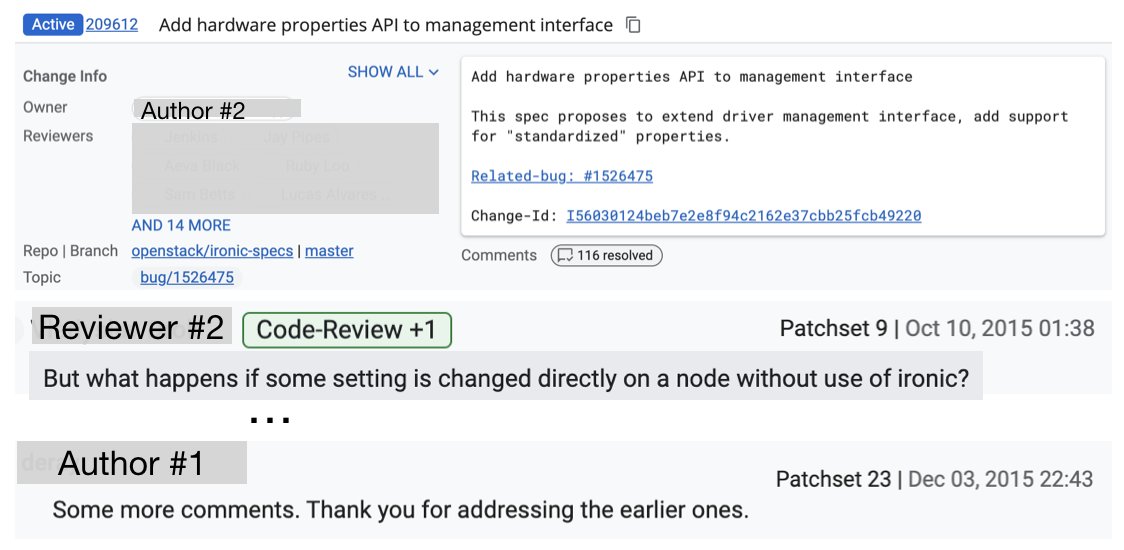}}
    \caption{A real example from OpenStack that illustrates the cross-patch collaborations after the patch linkage is posted. Note that to avoid ethical issues, the real developer names are anonymous.}
    \label{fig:illustration2}
\end{figure*}

Figure \ref{fig:illustration} illustrates a motivating scenario where collaboration occurs after the patch linkage.
As shown in the figure, a reviewer \textit{Pink} in Patch A posted a patch link to Patch B in the review discussion.
In this patch linkage, we consider Patch A as a source patch and Patch B as a target patch.
After the patch link is posted, a developer \textit{Green} who participated in the Patch A discussion votes and leaves review comments in Patch B.
At the same time, a developer \textit{Blue} who participated in the Patch B discussion before the linking time could also provide comments in Patch A discussion.
We consider either of these two cases as a collaboration occurrence.

In a realistic scenario (i.e., review at \url{https://review.openstack.org/#/c/211019}), as shown in Figure~\ref{fig:illustration2}, we observed that a reviewer (Reviewer \#1) posted a comment with a collaboration request to the patch author (Author \#1):
\begin{quote}
    \textit{`Could you please sync your efforts with another patch [https://review.openstack.org/\#/c\\/209612/]?'}
\end{quote}
After the patch link is posted, we observe that the author (Author \#1) and one of the reviewers (Reviewer \#2) from Patch \texttt{211019}, who were not involved in Patch \texttt{209612} before,  made the specific review comments related to the code changes in Patch \texttt{209612}.
Inspired by the realistic scenario, we hypothesize that there exist collaborations across patches (we called \textbf{cross-patch collaborations}, henceforth) after the patch linkage.

In this work, we conduct an empirical study of 368 patch linkages from a total of 8,612 linked patches to better understand the intentions of the patch linkage (e.g., requesting a collaboration) and statistically analyze to what extent collaboration will occur after the patch linkage.
Specifically, we investigate how different kinds of linkage sharing lead to collaboration opportunities and characterize the contribution kinds that follow after the link is identified.
Thus, three research questions are formulated to guide our study:
\begin{itemize}
\item \textbf{\RQOne}\\
\textit{\uline{Motivation.}}
Although Hirao et al.~\cite{hirao2019fse} have shown that patch linkage is mainly for team awareness (i.e., indicating dependency, providing broader context, and pointing out an alternative solution), we hypothesize that patch linkage could have an association with the developer collaboration across the patches. 
Thus, we would like to first understand how often developers request collaborations accompanied with shared patch linkages.
\item \textbf{\RqTwo}\\
\textit{\uline{Motivation.}}
Prior work sheds light that patch linkage can increase awareness~\cite{hirao2019fse}.
Yet, little is known about whether developers are likely to contribute to another via the patch linkage. 
To better understand this, we investigate to what degree collaboration will occur after a patch
link is posted.
\item \textbf{\RqThree}\\
\textit{\uline{Motivation.}}
To gain an in-depth insight, we would like to understand what kinds of collaborations developers do after the awareness of patch linkages.
Answering this question would help researchers and practitioners better understand the role of patch linkages.
\end{itemize}

The empirical results lead us to conclude that the patch linkage requesting collaboration is relatively less frequent. 
In addition, the delay exists before a patch linkage is posted (RQ1).
We observe that a cross-patch collaboration is more likely to occur when the intention of a request for collaboration is accompanied with the patch linkage (RQ2).
Specially, four kinds of collaboration activities are classified, including voting, writing specific and general comments, and a revision of linked patches (RQ3).

The remainder of this paper is organized as follows. 
Section 2 describes the empirical study design, including data preparation and approaches for each RQ.
Section 3 presents the results of our empirical study, while Section 4 discusses our findings and challenge. 
Section 5 discloses the threats to validity.
Section 6 discusses the related work regarding link sharing and reviewer participation in code reviews.
Finally, we conclude the paper in Section 7.

\section{Empirical Study Design}
\subsection {Data Collection}

In this study, we use OpenStack as a case ecosystem.
OpenStack is an open-source software ecosystem where many well-known organizations and companies, e.g., IBM, VMware, and NEC, collaboratively develop a platform for cloud computing. 
OpenStack actively performs code reviews through Gerrit, a tool-based code review tool, and is widely studied in the prior work~\cite{WANG_emse, chouchen2021anti, p65}.

\textbf{\textit{Clean Dataset.}} For our experiments, we used the OpenStack review dataset provided by Thongtanunam and Hassan~\cite{pick_tse}.
The dataset includes 58,212 patches dated from November 2011 to July 2019.
Since we focus on the collaboration and contributions done by patch authors or reviewers, we exclude the comments that are posted by automated tools in the discussion threads. To do so, we refer to the documentation of the studied system\footnote[2]{https://docs.openstack.org/infra/manual/developers.html} to identify the automated tools that are integrated with the code review tools.
Specifically, we use the list of the automated tools that is provided in the work of Thongtanunam et al. \cite{pick_tse}

\begin{table}[]
\caption{Summary of dataset used in the study.}
\label{tab:dataset}
\resizebox{0.5\textwidth}{!}{%
\begin{tabular}{@{}lrrr@{}}
\toprule
Ecosystem & Time Window & Linkage Dataset & Sample Dataset \\ \midrule
OpenStack & 2011.11 - 2019.7 & 8,612 & 368 \\
\bottomrule
\end{tabular}%
}
\end{table}

\textbf{\textit{Extract Patch Linkage.}} To identify the patch links, similar to prior work~\cite{WANG_emse}, we applied the regular expression to search all messages in the review discussions that include a patch URL in the following format: \textit{https?://review.openstack$\mid$opendev.org/\#/c/[1-9]+[0-9]*}.
A total of 8,944 pairs of patches are retrieved.
Then we exclude the case where the source and target patches are the same.
In our study, we keep the cases where (i) the patch linkages are written by the same patch authors and (ii) the patch authors post links by themselves, as we assume that collaboration could occur between the reviewers of both patches.
Finally, we obtain 8,612 pairs of patches that met our experiment criteria.
Table 1 shows the summary of the dataset used in the study.

\begin{table*}[t]
\tabcolsep=0.3cm
\centering
\caption{(RQ1) The prevalence of linkage kinds and their timing nature. Requesting collaboration accompanied with links is less common.}
    \label{table:prevalence}
\resizebox{\textwidth}{!}{
\begin{tabular}{llrrrrrrrr}
\toprule
\multirow{2}{*}{\textbf{Linkage Kind}}           & \multirow{2}{*}{\textbf{Count}} & \multicolumn{4}{c}{\textbf{Patch-linked Time (\# days)}} & \multicolumn{4}{c}{\textbf{Patch-closed Time (\# days)}} \\
                                     &                        & \textbf{1st Qu.}     & \textbf{Median}     & \textbf{Mean}    & \textbf{3rd Qu.}    & \textbf{1st Qu.}     & \textbf{Median}     & \textbf{Mean}    & \textbf{3rd Qu.}    \\ \midrule
Requesting collaboration             & {$\;\,$57  \mybar{0.3}}                     & 1.2         & 14.1       & 48.8    & 56.5       & 3.3         & 20.6       & 92.1    & 67.2       \\
Sharing information                  & 211 \mybar{1.14}                    & 0.9         & 11.6       & 38.3    & 48.2       & 1.0         & 10.8       & 55.1    & 49.2       \\
Pointing out an alternative solution & 100 \mybar{0.54}                   & 0.4         & 4.0        & 31.7    & 32.8       & 0.0         & 0.9        & 31.2    & 19.2       \\ \bottomrule
\end{tabular}}
\end{table*}

\subsection{RQ1 Analysis}
To answer \RQOne, we investigate the intention of posting patch linkages in the aspect of collaborations.
In addition, we conduct a statistical analysis to investigate the timeline of patch linkages (e.g., when the linkage is posted and how long it takes the review to be completed after the linkage is posted). 
Such timeline analysis could highlight the necessity of tool support for in-time linkage recommendations.
Below, we describe these two analysis approaches in detail.

\textbf{Requesting collaboration.} We perform a manual analysis to investigate the intention behind the patch linkage.
More specifically, our analysis mainly focuses on how often the patches are linked to request collaboration. 
Below, we describe our manual coding based on a statistically representative sample of our patch linkage dataset:

\textit{Representative dataset construction.}
    As the full set of our constructed data is too large to manually examine their collaboration intention, we then draw a statistically representative sample.
    The calculation of
    statistically significant sample sizes based on population size, confidence interval, and confidence level is well established~\cite{sample_size}, with a confidence level of 95\% and a confidence interval of 5.
    In the end, we randomly sample 368 patch linkages.\footnote[3]{https://www.surveysystem.com/sscalc.htm}
    
    \textit{Manual coding.} 
    In this step, we classify whether the patch linkage is for requesting collaboration or not. Based on the finding of prior work~\cite{hirao2019fse}, patch linkage can be also for sharing information or pointing out an alternative solution. Hence, we classify the intention of patch linkages into three main kinds: 
    \begin{itemize}
        \item \textit{Requesting collaboration}: Patch linkage for requesting collaboration is the linkage where a developer (either a patch author or reviewer) posts a link with a message that explicitly requests other developers to collaborate in the target patch. 
        In this case, developers often write message which includes  words such as `help', `collaborate', `integrate' or `rebase on'.
        For example, \textit{``Patch Set 1: Code-Review-1
        Can we please rebase this on https://review.openstack.org/\#/c/93842/ that review ensures specific values is present in the string for the flag to be switched on. thanks, dims''}.
        \item \textit{Sharing information}: Patch linkage for sharing information is the linkage where a developer posts a link to increase team awareness (e.g., indicating patch dependency, providing broader context)
        \item \textit{Pointing out an alternative solution}: 
        Patch linkage for pointing out an alternative solution is the linkage where a developer posts a link to mention that the target patch attempts to explicitly address the same or similar objective as the source patch.
  \end{itemize}
  To classify the patch linkages into a category, we consider the whole textual message that comes with the link.
  In some cases, we also read the whole review discussion to understand the context.
  To test the comprehensive understanding of the constructed schema, we randomly select 30 samples from our representative dataset, and the three authors of this paper independently coded these samples.
  Among the three coders, we obtain a Kappa agreement score of 0.77 (i.e., substantial). The three coders then discussed the samples with inconsistent codes to reach a consensus.
  Encouraged by the promising Kappa agreement score, the remaining data was then coded by one coder.

\textbf{Timeline of patch linkage.}
To understand the timeline of patch linkage, we measure patch-linked time and patch-closed time.
The patch-linked time is the duration from when reviews start on a patch to the time when the patch link is posted into the review discussion.
The patch-closed time is the duration from when a patch link is posted to the time when the review is closed.
We assume that patch-linked time and patch-closed time significantly differ among linkage categories (i.e., requesting collaboration, sharing information, and pointing out an alternative solution), and specifically a relatively longer time is likely taken for the linkage of requesting collaboration to be linked and closed.
Then, we perform a statistical analysis to examine our assumption.
To do so, we use a Kruskal-Wallis test, i.e., a non-parametric test, to compute the statistical significance.

\subsection{RQ2 Analysis}
To answer \RqTwo, we investigate how frequently the collaboration will occur after the patch linkage.
Below, we describe how we measure the collaboration occurrence.

\textbf{Collaboration occurrence.} We analyze the set of additional developers who newly join and contribute to the patch after the patch link is posted.
We consider both directions of collaboration, i.e., developers who participate in the source patch contribute to the target patch (Source $\rightarrow$ Target) and developers who participate in the target patch contribute to the source patch (Source $\leftarrow$ Target).

To identify the additional developers and direction of collaboration, we first identify the set of developers who contribute (e.g., providing a comment, voting) to the source patch before the patch link is posted (S) and the set of other developers who *only* contribute after the link is posted (S').
Note that S includes the developer who posted the patch link.
Similarly, we identify the set of developers who contribute to the target patch based on the time point when the patch link is posted (T and T').
Then, we identify the set of developers in the source patch who contribute to the target patch after the patch link is posted (i.e., Source $\rightarrow$ Target = S $\cap$ T') and the number of developers in the target patch who contribute to the source patch after the patch link is posted (i.e., Source $\leftarrow$ Target = T $\cap$ S').

For example, in Figure \ref{fig:illustration}, we will identify the following sets of developers: S = \{Green, Pink\}, S' = \{Blue\}, T = \{Blue, Orange\}, and T' = \{Green\}.
Therefore, in this example, the developer \textit{Green} is considered as the one who is from the source patch and contributes to the target patch.
Similarly, the developer \textit{Blue} is considered as the one who is from the target patch and contributes to the source patch.
Note that since we will analyze the collaboration occurrence across the three link kinds, we perform this analysis based on the labeled 368 patch linkages.

\subsection{RQ3 Analysis}
To answer \RqThree, we conduct a semi-automatic analysis to further investigate the kinds of collaboration activities of developers who newly join and contribute to the patch. 
Below, we describe the approach to identify collaboration activities.

\textbf{Collaborative contribution kinds.}
In addition to the occurrence analysis, we examine what collaborative contributions were made by the additional developers (i.e., S $\cap$ T' and S' $\cap$ T).
In this work, based on an open discussion with ten random samples and the OpenStack documentation by the first three authors of this paper, we focus on four kinds of contributions: 1) \textit{Vote}, 2) \textit{Specific Comments}, 3) \textit{General Comments}, and 4) \textit{Revise}.
Table~\ref{table:contribution} describes the definition of four contribution kinds.

\begin{table}[b]
\centering
\caption{The definition of contribution kinds and their distribution across the link kinds. Note that one review message can be labeled with more than one contribution kind.}
\label{table:contribution}
\begin{tabular}{lp{4.4cm}}
\toprule
\textbf{Contribution Kind}                  & \textbf{Definition}  \\ \hline
Vote             & Collaborator votes whether to merge or abandon the patch, i.e., ``Code-Review +1''.\\ \midrule
Specific Comments & Collaborator posts a comment that is directly related to patch change, i.e.,  typically an inline comment to reference a line of code in the patch.        \\ \midrule
General Comments  & Collaborator posts a generic comment that does not directly relate to or reference any line of code in the patch.     \\ \midrule
Revise           & Collaborator uploads revised patches,  i.e., ``Uploaded patch set 3''.      \\ 
                                   \bottomrule
\end{tabular}
\end{table}

To identify the contribution kinds, we extract the contribution information recorded in the review message, i.e., 1,898 contributions are retrieved from the 368 labeled patch linkages.
We classify the collaborative contribution kinds in two rounds.
In the first round,  we use a regular expression to automatically identify each kind of contribution.
Specifically, we use the expressions (?:.*(Workflow[\textbackslash+|\textbackslash-][0-9]|Code-Review[\textbackslash+|\textbackslash-][0-9]).*),  (?:\textbackslash(.*?comment.*?\textbackslash)), and (?:uploaded patch set) to identify \textit{Vote}, \textit{Specific Comments}, and \textit{Revise}, respectively.
The rest of them are classified as \textit{General Comments}.
In the second round, we manually validate the identified candidates to reduce the potential threats caused by false positives.
In addition, we highlight those general comments that are not trivial.
For instance, a not trivial general comment is left with \textit{``Patch Set 2: Code-Review-1
i think you should update this file 
https://github.com/openstack/neutron/blob/master/do\\c/requirements.txt} because after the new PTI, doc requirements are moved here.''

\begin{table*}[t]
\centering
\caption{(RQ2) Collaboration occurrence between the source patch and target patch. Collaboration is more likely to occur when the request is provided.}
\label{table:occurrence}
\resizebox{.9\textwidth}{!}{
\begin{tabular}{llr}
\toprule
\textbf{Collaboration Direction}                               & \textbf{Linkage Kind}                     & \textbf{Occurrence Percent}   \\ \midrule
\multirow{4}{*}{Source $\rightarrow$ Target} & Requesting collaboration        & \blackwhitebar{0.72}\\
                               & Sharing information            & \blackwhitebar{0.57}              \\
                               & Pointing out an alternative solution & \blackwhitebar{0.47}                   \\
                               & Average & \blackwhitebar{0.57}\\ \midrule
\multirow{4}{*}{Source $\leftarrow$ Target} & Requesting collaboration            & \blackwhitebar{0.62}                    \\
                               & Sharing information        & \blackwhitebar{0.49}                   \\
                               & Pointing out an alternative solution & \blackwhitebar{0.34}                    \\
                               & Average & \blackwhitebar{0.47}\\ \bottomrule
\end{tabular}}
\end{table*}

\section{Empirical Results}
In this section, we present the results for each of our research questions.

\subsection{RQ1: To what degree do developers request collaborations when posting patch linkages?}

\textit{Results:} We observe two main findings.
\ul{First, patch linkage for requesting collaboration is relatively less frequent than others.}
Table~\ref{table:prevalence} shows that only 57 patch linkages (15\%) where developers post a patch link with an explicit request for collaboration.
Most patch linkages (i.e., 211 patch linkages) are posted for sharing information such as patch dependency and broader context, while the other 100 patch linkages are for pointing out an alternative solution.

\ul{Second, we observe around 4 to 14 days (median) before a review member posts a patch linkage.}
Regarding the patch-linked time,
we find that it takes a relatively long time for review teams to post patch linkage.
The median of 14.1, 11.6, and 4.0 days are taken for requesting collaboration, sharing information, and pointing out an alternative solution, respectively, as shown in Table~\ref{table:prevalence}. 
Related to the patch-closed time, we find that the patch with the linkage indicating an alternative solution is more likely to be closed quicker than the other categories.
Interestingly, we find that the linkage for requesting collaboration takes a longer time to be closed compared with other categories. 
It is also important to note that we do not aim to draw a causal relationship, but only observe a trend.
Several confounding factors may also play a role, such as the patch size, the patch complexity, and the extent of the change impacts made by the patch.
For instance, if a patch is complex to be understood, it may require more collaboration and further take a longer review time.
The Kruskal-Wallis test confirms that there is a significant difference ($p$-value $<$ 0.001) in the patch-linked time and patch-closed time among different linkage kinds.
Moreover, our assumption that a relatively longer long time is taken for the linkage of requesting collaboration to be linked and closed is established.

\begin{tcolorbox}
\textbf{Takeaway I:}
Patch linkage that requests collaboration is relatively less prevalent.
Furthermore, we observe a delay from 4 to 14 days before a patch linkage is posted.
\end{tcolorbox}

\subsection{RQ2: How likely will collaborations occur after patch linkages are posted?}

\textit{Results:} \ul{Patch linkage with requesting collaboration has a relatively higher percentage of collaboration than the other two kinds.}
Table~\ref{table:occurrence} shows the percentage of patch linkages that have at least one developer from the source patch who contributes to the target patch or vice versa.
We find that on average, 72\% of the patch linkages for requesting collaboration have at least one developer from the source patch who contributes to the target patch (Source $\rightarrow$ Target).
Similarly, 62\% of the patch linkages for requesting collaboration have at least one developer from the target patch contributing to the source patch (Source $\leftarrow$ Target).
On the other hand, the percentages of collaboration in the other two link kinds are relatively lower than the percentage of the patch linkages for requesting collaboration (i.e., 34\%--57\%).
\begin{tcolorbox}
\textbf{Takeaway II:}
A cross-patch collaboration is more likely to occur when the patch linkage comment is accompanied with a request for collaboration.
\end{tcolorbox}

\begin{table*}[t]
\centering
\caption{(RQ3) The frequency of four kinds of cross-patch collaborations. Vote and general comments are more common contributions.}
\label{table:contribution2}
\resizebox{.8\textwidth}{!}{
\begin{tabular}{lll}
\toprule
\textbf{Contribution after Linkage}                                                                                                                                                                                & \textbf{Link Kind}                            & \textbf{Percent} \\ \hline
\multirow{3}{*}{Vote}                                                               & Requesting collaboration             & \blackwhitebar{0.56}            \\
                                   &     Sharing information   &       \blackwhitebar{0.59}     \\
                                   &      Pointing out an alternative solution &     \blackwhitebar{0.61}       \\ \midrule
\multirow{3}{*}{Specific Comments} &  Requesting collaboration             &   \blackwhitebar{0.29}          \\
                                   &     Sharing information                  &      \blackwhitebar{0.37}       \\
                                   &    Pointing out an alternative solution &    \blackwhitebar{0.31}         \\ \midrule
\multirow{3}{*}{General Comments}  & Requesting collaboration             &      \blackwhitebar{0.43}       \\
                                   &                                                                                                                                                                            Sharing information                  &      \blackwhitebar{0.47}       \\
                                   &                                                                                                                                                                            Pointing out an alternative solution &      \blackwhitebar{0.52}       \\ \midrule
\multirow{3}{*}{Revise}            &  Requesting collaboration             &     \blackwhitebar{0.15}        \\
                                   &                                                                                                                                                                            Sharing information                  &       \blackwhitebar{0.09}      \\
                                   &                                                                                                                                                                            Pointing out an alternative solution &     \blackwhitebar{0.05}        \\ 
                                   \bottomrule
\end{tabular}}
\end{table*}

\subsection{RQ3: What are the kinds of cross-patch collaboration activities?}
\textit{Results:} \ul{Among four kinds of collaboration activities, vote is the most frequent contribution kind.}
Table~\ref{table:contribution2} shows the distribution of contribution kinds across the link kinds.
We find that among four contribution kinds, \textit{Vote} is the most common kind (i.e., 56\% for requesting collaboration, 59\% for sharing information, and 61\% for pointing out an alternative solution).
The following common contribution kind is \textit{General Comments}.
Based on our manual validation, we observe that 19.7\% of these comments are left with not trivial information.
For instance, one comment provides advice to fix up the eventlet change, i.e., \textit{``Patch Set 4:
OK, fix up the docstring on run\_vios\_command\_as\_root and I think the commit message should mention the eventlet change, and then I'm +1.''}.
Interestingly, we find that \textit{Revise} contribution kind is relatively more frequent in the patch linkage for requesting collaboration (15\%) than the other two link kinds (9\% and 5\%, respectively).
This result suggests that the patch linkage for requesting collaboration is more likely to trigger the collaborative activity related to the patch quality (i.e., where the collaborator uploads revised patches).

\begin{tcolorbox}
\textbf{Takeaway III:}
The cross-patch collaboration via the patch linkage includes voting, writing specific and general comments, and a revision of patches. 
Furthermore, voting is the most common collaboration kind, i.e., 57\% being identified on average.
\end{tcolorbox}

\section{Threats to Validity}
We now discuss the threats to the validity of our empirical study.

\textit{External Validity.} 
External validity is concerned with our ability to generalize based on our results.
Our study only focuses on the OSS ecosystem (i.e., where multiple projects develop software collaboratively) using a tool-based code review.
We understand that there are not many multi-project review ecosystems similar to OpenStack.
However, as open source
adoption has grown significantly in the last decade, and numerous companies have built business models around OSS ecosystems~\cite{Zhang2020HowDC}, we believe it is important to study the OSS ecosystem.

\textit{Internal Validity.} Internal validity is the approximate
truth about inferences regarding cause-effect or causal relationships.
In our empirical study,  we employ manual analysis for classifying linkage kinds.
The label might be miscoded due to the subjective nature of understanding. To eliminate such a threat, we use the Kappa agreement to measure inter-rater reliability.
Only until the Kappa score reaches more than 0.7 (i.e., the score in classifying intentions is 0.77, indicating substantial agreement), we were able to complete the rest of the samples.
Another threat may occur in the choice of selecting statistical test techniques.
To address the statistical significance of the timeline, we apply the Kruskal-Wallis test, a non-parametric test.
While, we are confident with this test, which is widely used in the previous work~\cite{wang_IST}.

\textit{Construct Validity.} Construct validity is concerned with
the degree to which our measurements capture what we
aim to study.
This threat potentially occurs in the extraction of identifying patch linkages.
In our study, we only extract patch links from the general code review discussions, however, the patch links may also appear in inline review discussions.
We believe this will not affect our observations, since we use qualitative analysis to investigate the patch links in this study.
Another threat could be concerning the validity of the categorization in RQ1. We classify three intentions behind the patch linkage, i.e., requesting collaboration, sharing information, and pointing out an alternative solution. Sharing information may be a part of collaboration. 
To ensure that there is no orthogonality between them, we only classify the comment that provides actionable collaboration intention (e.g., `help', `collaborate', and so on) as requesting collaboration; otherwise, we classify it as sharing information.

\section{Challenges and Opportunities}
\label{conclusion}
We now discuss our empirical findings and challenges, as well we provide several possible opportunities to guide future research. 

The empirical results show the potential for this new kind of collaboration that is triggered by a patch linkage. 
Hence, the study calls for new avenues for research into this kind of collaboration. 
In fact, we show that cross-patch collaborative contributions via the patch linkage are non-trivial, with key contributions like voting which affects the review outcome of the target patch, or revising which improves the patch.
In terms of the timeline of patch linkage, our empirical study provides evidence that to be aware of the existence of patch linkage takes a relatively longer time.

There are still open challenges that remain.
For instance, the current approach has the threat to include collaborations that may have not been triggered by the patch linkage.
Hence, future work needs to address the soundness of our approach. 
Another challenge may include capturing cross-patch collaborations that do not have patch linkage. 
This can also be addressed in a bigger study.
Furthermore, we would need a developer study to validate the practical implications of the study. 

Our work lays out future opportunities for directions on how patch linkage sharing can lead to these new kinds of collaboration.
We highlight three below to name a few:
\begin{itemize}
    \item \textit{Identify heuristics and the information required for a reviewer to contribute to a linked patch}. To gain more practical insights, a survey or interview of the reviewer who posts the link could reveal collaboration barriers and opportunities
    \item \textit{Investigate the impact of the collaboration on patch quality and code review quality.} To further understand the impact of the collaboration, one promising direction is to explore if the patch involved with contribution via the linkage is likely to decrease the probability of defects.
    \item \textit{Automatic recovery of links (especially for Duplicate/Alternative Solution Detection)}. Provide tool support to early detect or recommend patches to reduce the time taken to identify the link, especially since we find that pointing out an alternative solution earlier leads to a shorter review time compared to the other link kinds.
\end{itemize}

\section{Related Work}
In this section, we position our work with literature reviews in terms of the practice of link sharing in software engineering and the reviewer participation in the context of code review settings.

\subsection{The Practice of Link Sharing}
Link sharing has become a popular activity in software engineering, which enables developers to share knowledge and mitigate potential issues.
The value of link sharing has been commonly addressed in question-and-answer forums like Stack Overflow and tool-based code reviews like GitHub and Gerrit.
Goemz et al.~\cite{gomez_2013} reported that link sharing is a significant phenomenon on
Stack Overflow, referring readers to software development innovations like libraries and tools.
Ye et al.~\cite{stack_emse_2017} generated the structural and dynamic properties of the emergent knowledge network, using shared URLs in Stack Overflow.
With the popularity of GitHub, Hata et al.~\cite{Hata:ICSE19} found that 9.6 million links exist in source code comments across 25,925 repositories. 
Within Gerrit based reviews, Wang et al.~\cite{WANG_emse} observed seven intentions behind link sharing and their developer survey results suggested that link sharing is useful.
At the same time, Hirao et al.~\cite{hirao2019fse} categorized five kinds of review linkage, such as patch dependency, broader context, and alternative solution.
To aid such practice, Wang et al.~\cite{wang_IST} proposed a linkage detection using textual contents and file location.

Our work expanded upon the work of Wang et al.~\cite{WANG_emse} and Hirao et al.~\cite{hirao2019fse} to investigate the aspect of collaboration across patch linkages.

\subsection{Reviewer Participation}
The reviewer participation becomes one of the main challenges in the tool-based code review process, since unlike formal code review, reviewers
can decide whether or not to participate in a review.
A large body of studies has found that reviewer participation is associated with software quality and code review time~\cite{p28,pj3,chouchen2021anti}. 
For instance, Kononenko et al.~\cite{p09} observed that the number of invited
reviewers have a statistically significant impact on review bugginess.
Moreover, Ruangwan et al.~\cite{pj28} reported that human factors play an important role in predicting whether or not an invited reviewer will participate in a review.
To relieve the challenges of reviewer participation, many reviewer recommendation systems have been proposed.
Thongtanunam et al.~\cite{p65} introduced REVFINDER, a file location-based code-reviewer recommendation approach.
Xia et al.~\cite{p34} put textual information and file location analyses together to recommend reviewers more accurately.
Hannebauer et al.~\cite{p48} recommended code reviewers based on their expertise.
Most recently, Al-Zubaidi et al.~\cite{al2020workload} developed a novel approach that
leverages a multi-objective meta-heuristic algorithm to search for reviewers guided by two objectives.

Similarly, the ultimate goal of our study is to improve the reviewer participation by understanding the developer collaboration activities across patch linkages.

\section{Conclusion and Future Work}
The growing number of reviews in open source projects poses a new challenge for collaboration during the review process and development tasks.
In this paper, we perform an empirical study on OpenStack to investigate the cross-collaborations via patch linkages.
Our results show that requesting collaboration accompanied with shared patch links is less common, while cross-patch collaboration is more likely to occur once the request is provided.
Moreover, four kinds of collaboration activities are classified and the results suggest that cross-collaborations are not trivial.
Future research directions include the causality analysis between patch linkage and collaboration, perceptions and collaboration barriers from real developers, and tool development for link recovery.

\section*{Acknowledgment}
This work has been supported by JSPS KAKENHI Grant Numbers JP20K19774, and JP20H05706.
P. Thongtanunam
was supported by the Australian Research Council’s Discovery
Early Career Researcher Award (DECRA) funding scheme (DE210101091).

\bibliographystyle{ieicetr}
\bibliography{reference}

\begin{thebibliography}{10}

\bibitem{WANG2021111009}
D.~Wang, Y.~Ueda, R.G. Kula, T.~Ishio, and K.~Matsumoto, ``Can we benchmark
  code review studies? a systematic mapping study of methodology, dataset, and
  metric,'' J. Syst. Softw, vol.180, p.111009, 2021.

\bibitem{Google_2018}
C.~Sadowski, E.~Söderberg, L.~Church, M.~Sipko, and A.~Bacchelli, ``{Modern
  Code Review: A Case Study at Google},'' Proc. 39th Int. Conf. on Software
  Engineering: Software Engineering in Practice Track, p.181–190, 2018.

\bibitem{FSE2013_Rigby}
P.C. Rigby and C.~Bird, ``{Convergent Contemporary Software Peer Review
  Practices},'' Proc. 9th Joint Meeting on Foundations of Software Engineering,
  p.202–212, 2013.

\bibitem{ICSME18_pull}
X.~Zhang, Y.~Chen, W.Z. Yongfeng~Gu, X.~Xie, X.~Jia, and J.~Xuan, ``{How do
  Multiple Pull Requests Change the Same Code: A Study of Competing Pull
  Requests in GitHub},'' Proc. 34th Int. Conf. on Software Maintenance and
  Evolution, pp.228--239, 2018.

\bibitem{ebert2019confusion}
F.~Ebert, F.~Castor, N.~Novielli, and A.~Serebrenik, ``Confusion in code
  reviews: Reasons, impacts, and coping strategies,'' Int. Conf. on Software
  Analysis, Evolution and Reengineering, pp.49--60, 2019.

\bibitem{WANG_emse}
D.~Wang, T.~Xiao, P.~Thongtanunam, R.G. Kula, and K.~Matsumoto, ``Understanding
  shared links and their intentions to meet information needs in modern code
  review,'' Empir. Softw. Eng., vol.26, no.5, p.96, 2021.

\bibitem{hirao2019fse}
T.~Hirao, S.~McIntosh, A.~Ihara, and K.~Matsumoto, ``{The Review Linkage Graph
  for Code Review Analytics: A Recovery Approach and Empirical Study},'' Proc.
  Int. Symp. on the Foundations of Software Engineering, p.578–589, 2019.

\bibitem{chouchen2021anti}
M.~Chouchen, A.~Ouni, R.G. Kula, D.~Wang, P.~Thongtanunam, M.W. Mkaouer, and
  K.~Matsumoto, ``Anti-patterns in modern code review: Symptoms and
  prevalence,'' IEEE Int. Conf. on Software Analysis, Evolution and
  Reengineering (SANER), pp.531--535, IEEE, 2021.

\bibitem{p65}
P.~{Thongtanunam}, C.~{Tantithamthavorn}, R.G. {Kula}, N.~{Yoshida}, H.~{Iida},
  and K.~{Matsumoto}, ``Who should review my code? a file location-based
  code-reviewer recommendation approach for modern code review,'' IEEE 22nd
  Int. Conf. on Software Analysis, Evolution, and Reengineering (SANER),
  pp.141--150, 2015.

\bibitem{pick_tse}
P.~{Thongtanunam} and A.E. {Hassan}, ``Review dynamics and their impact on
  software quality,'' IEEE Trans.\ Software Eng., pp.1--1, 2020.

\bibitem{sample_size}
R.V. Krejcie and D.W. Morgan, ``Determining sample size for research
  activities,'' Educational and Psychological Measurement, 1970.

\bibitem{Zhang2020HowDC}
Y.~xia Zhang, M.~Zhou, K.J. Stol, J.~yu~Wu, and Z.~Jin, ``How do companies
  collaborate in open source ecosystems? an empirical study of openstack,''
  IEEE/ACM 42nd Int. Conf. on Software Engineering, p.1196–1208, 2020.

\bibitem{wang_IST}
D.~Wang, R.G. Kula, T.~Ishio, and K.~Matsumoto, ``Automatic patch linkage
  detection in code review using textual content and file location features,''
  Inf. Softw. Technol., vol.139, no.C, nov\ 2021.

\bibitem{gomez_2013}
C.~Gomez, B.~Cleary, and L.~Singer, ``A study of innovation diffusion through
  link sharing on stack overflow,'' Int. Working Conf. on Mining Software
  Repositories, pp.81--84, 2013.

\bibitem{stack_emse_2017}
D.~Ye, Z.~Xing, and N.~Kapre, ``The structure and dynamics of knowledge network
  in domain-specific q\&a sites: A case study of stack overflow,'' Empir.
  Softw. Eng., p.375–406, 2017.

\bibitem{Hata:ICSE19}
H.~Hata, C.~Treude, R.G. Kula, and T.~Ishio, ``{9.6 Million Links in Source
  Code Comments: Purpose, Evolution, and Decay},'' Proc. Int. Conf. on Software
  Engineering, 2019.

\bibitem{p28}
S.~McIntosh, Y.~Kamei, B.~Adams, and A.E. Hassan, ``The impact of code review
  coverage and code review participation on software quality: A case study of
  the qt, vtk, and itk projects,'' Proc. 11th Working Conf. on Mining Software
  Repositories, MSR 2014, pp.192--201, 2014.

\bibitem{pj3}
P.~Thongtanunam, S.~Mcintosh, A.E. Hassan, and H.~Iida, ``Review participation
  in modern code review,'' Empir. Softw. Eng., pp.768--817, 2017.

\bibitem{p09}
O.~{Kononenko}, O.~{Baysal}, and M.W. {Godfrey}, ``Code review quality: How
  developers see it,'' 38th Int. Conf. on Software Engineering (ICSE),
  pp.1028--1038, 2016.

\bibitem{pj28}
S.~Ruangwan, P.~Thongtanunam, A.~Ihara, and K.~Matsumoto, ``The impact of human
  factors on the participation decision of reviewers in modern code review,''
  Empir. Softw. Eng., p.973–1016, 2018.

\bibitem{p34}
X.~Xia, D.~Lo, X.~Wang, and X.~Yang, ``Who should review this change?: Putting
  text and file location analyses together for more accurate recommendations,''
  IEEE Int. Conf. on Software Maintenance and Evolution (ICSME), pp.261--270,
  09\ 2015.

\bibitem{p48}
C.~Hannebauer, M.~Patalas, S.~St\"{u}nkel, and V.~Gruhn, ``Automatically
  recommending code reviewers based on their expertise: An empirical
  comparison,'' Proc. Int. Conf. on Automated Software Engineering, pp.99--110,
  2016.

\bibitem{al2020workload}
W.H.A. Al-Zubaidi, P.~Thongtanunam, H.K. Dam, C.~Tantithamthavorn, and
  A.~Ghose, ``Workload-aware reviewer recommendation using a multi-objective
  search-based approach,'' Proc. Int. Conf. on Predictive Models and Data
  Analytics in Software Engineering, pp.21--30, 2020.

\end{thebibliography}

\newpage
\profile[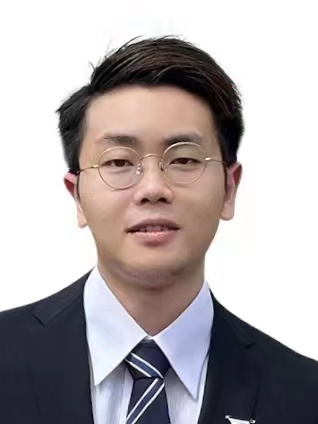]{Dong Wang}{is currently an assistant professor at Kyushu University. He received PhD from Nara Institute of Science and Technology, Japan. He is a member of the IEEE Computer Society. His research interests include mining software repositories, empirical software engineering, and human aspects. More about his information is available online at \url{https://dong-w.github.io/}.}

\profile[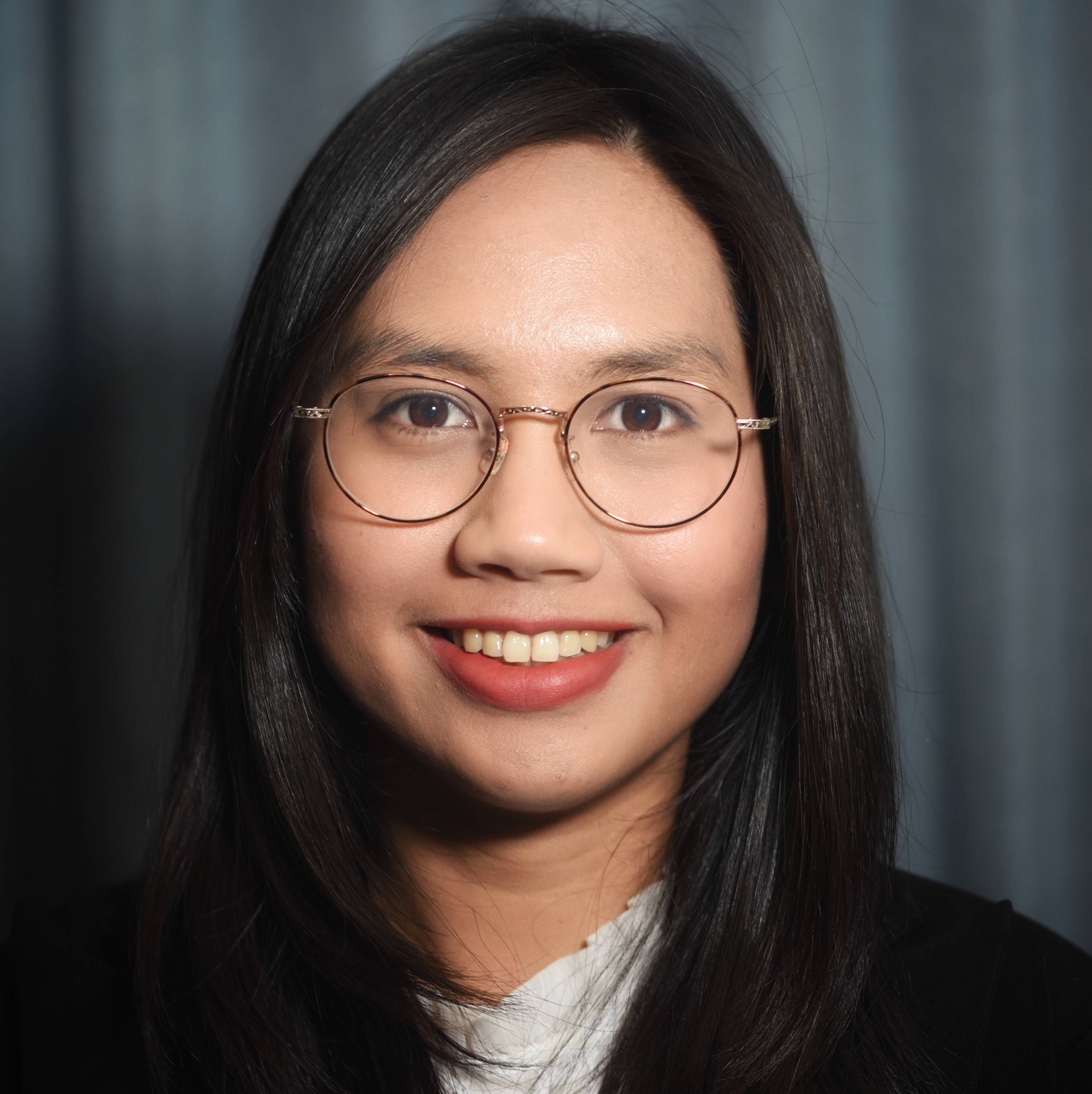]{Patanamon Thongtanunam}{is an ARC DECRA awardee and a senior lecturer at the School of Computing and Information System, the University of Melbourne, Australia. She received PhD from Nara Institute of Science and Technology, Japan. Her research interests include empirical software engineering, mining software repositories, software quality, and human aspect. Her research has been published at top-tier software engineering venues like ICSE, TSE, and EMSE. More about Patanamon and her work is available online at \url{http://patanamon.com}.}

\profile[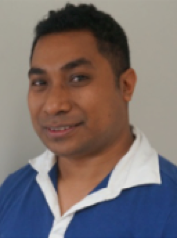]{Raula Gaikovina Kula}{is an assistant professor at Nara Institute of Science and Technology. In 2013, he graduated with a PhD. from Nara Institute of Science and Technology, Japan. He is currently an active member of the IEEE Computer Society and ACM. His research interests include repository mining, code review, software libraries and visualizations.}

\profile[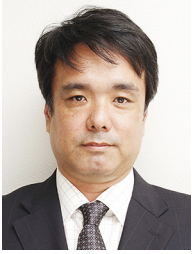]{Kenichi Matsumoto}{received the B.E., M.E., and PhD degrees in Engineering from Osaka University, Japan, in 1985, 1987, 1990, respectively. Dr. Matsumoto is currently a professor in the Graduate School of Information Science at Nara Institute Science and Technology, Japan. His research interests include software measurement and software process. He is a senior member of the IEEE and a member of the IPSJ and SPM.}

\end{document}